\documentclass[%
 aip,
 amsmath,amssymb,
 reprint,%
]{revtex4-2}

\usepackage{graphicx}
\usepackage{blindtext}
\usepackage{color}
\usepackage{dcolumn}

\begin{document}

\title{Flux ramp modulation based MHz frequency-division dc-SQUID multiplexer}

\author{Daniel~Richter}
\affiliation{Kirchhoff-Institute for Physics, Heidelberg University, Im Neuenheimer Feld 227, 69120 Heidelberg, Germany.}
\author{Ludwig~Hoibl}
\affiliation{Kirchhoff-Institute for Physics, Heidelberg University, Im Neuenheimer Feld 227, 69120 Heidelberg, Germany.}
\author{Thomas~Wolber}
\affiliation{Institute for Data Processing and Electronics, Karlsruher Institute of Technology, Hermann-von-Helmholtz-Platz 1, 76344 Eggenstein-Leopoldshafen, Germany.}
\author{Nick~Karcher}
\affiliation{Institute for Data Processing and Electronics, Karlsruher Institute of Technology, Hermann-von-Helmholtz-Platz 1, 76344 Eggenstein-Leopoldshafen, Germany.}
\author{Andreas~Fleischmann}
\affiliation{Kirchhoff-Institute for Physics, Heidelberg University, Im Neuenheimer Feld 227, 69120 Heidelberg, Germany.}
\author{Christian~Enss}
\affiliation{Kirchhoff-Institute for Physics, Heidelberg University, Im Neuenheimer Feld 227, 69120 Heidelberg, Germany.}
\author{Marc~Weber}
\affiliation{Institute for Data Processing and Electronics, Karlsruher Institute of Technology, Hermann-von-Helmholtz-Platz 1, 76344 Eggenstein-Leopoldshafen, Germany.}
\author{Oliver~Sander}
\affiliation{Institute for Data Processing and Electronics, Karlsruher Institute of Technology, Hermann-von-Helmholtz-Platz 1, 76344 Eggenstein-Leopoldshafen, Germany.}
\author{Sebastian~Kempf}
\email[]{sebastian.kempf@kit.edu}
\affiliation{Kirchhoff-Institute for Physics, Heidelberg University, Im Neuenheimer Feld 227, 69120 Heidelberg, Germany.}
\affiliation{Institute for Micro- and Nanoelectronic Systems, Karlsruher Institute of Technology, Hertzstraße 16, 76187 Karlsruhe, Germany.}

\date{\today}

\begin{abstract}
We present a MHz frequency-division dc-SQUID multiplexer that is based on flux ramp modulation and a series array of $N$ identical current-sensing dc-SQUIDs with tightly coupled input coil. By running a periodic, sawtooth-shaped current signal through additional modulation coils being tightly, but non-uniformly coupled to the individual SQUIDs, the voltage drop across the array changes according to the sum of the flux-to-voltage characteristics of the individual SQUIDs within each cycle of the modulation signal. In this mode of operation, an input signal injected in the input coil of one of the SQUIDs and being quasi-static within a time frame adds a constant flux offset and leads to a phase shift of the associated SQUID characteristics. The latter is proportional to the input signal and can be inferred by channelizing and down-converting the sampled array output voltage. Using a prototype multiplexer as well as a custom readout electronics, we demonstrate the simultaneous readout of four signal sources with MHz bandwidth per channel.
\end{abstract}


\maketitle

Direct-current superconducting quantum interference devices (dc-SQUIDs) are presently one of the most sensitive wideband devices for measuring any physical quantity that can be naturally converted into magnetic flux. For this reason, dc-SQUIDs are nowadays routinely used for applications ranging from investigations of magnetic nanoparticles to diagnostics in health care or the exploration of mineral deposits \cite{Fag06}. The intrinsic compatibility with mK operation temperatures as well as the excellent noise performance make SQUIDs also key components for the readout of cryogenic particle detectors \cite{Ens05}.

The maturity of fabrication technology allows building SQUID systems with hundreds or thousands of identical sensors. Moreover, the size of present-day SQUID systems is not limited by fabrication technology but other system constraints such as cooling power or system complexity. This particularly applies to SQUID systems operating at mK-temperatures as used, for example, to read out cryogenic particle detectors. For this reason, multiplexing techniques are required to realize multi-channel SQUID systems providing ultra-low power dissipation at cryogenic temperatures, a readout bandwidth of several 10\,kHz up to some MHz, a large dynamic range  as well as a linear relation between the input and output signal.

Existing SQUID-based multiplexing techniques include time-division multiplexing \cite{RN469}, frequency-division multiplexing using MHz \cite{RN620} and GHz \cite{RN820} carriers, code-division multiplexing \cite{RN176} as well as hybrid multiplexing schemes \cite{RN725,RN731}. But despite of the great success and their numerous advantages, they suffer from minor drawbacks that practically challenge their application: MHz frequency-division multiplexers, for example, employ large on-chip passive filter circuits limiting the overall channel count per given chip area. In addition, parasitic impedances within the readout circuit as well as ac biasing of the sensors leads to severe effects that complicate the sensor readout\cite{RN902, RN903}. The frame rate of time-division SQUID multiplexers is too low to acquire wideband signals without signal deterioration. Though GHz frequency multiplexing can easily resolve these issues, it comes at the expense of an elaborated cryogenic microwave setup as well as a complex readout electronics.

\begin{figure*}
    \includegraphics[width=1.0\linewidth]{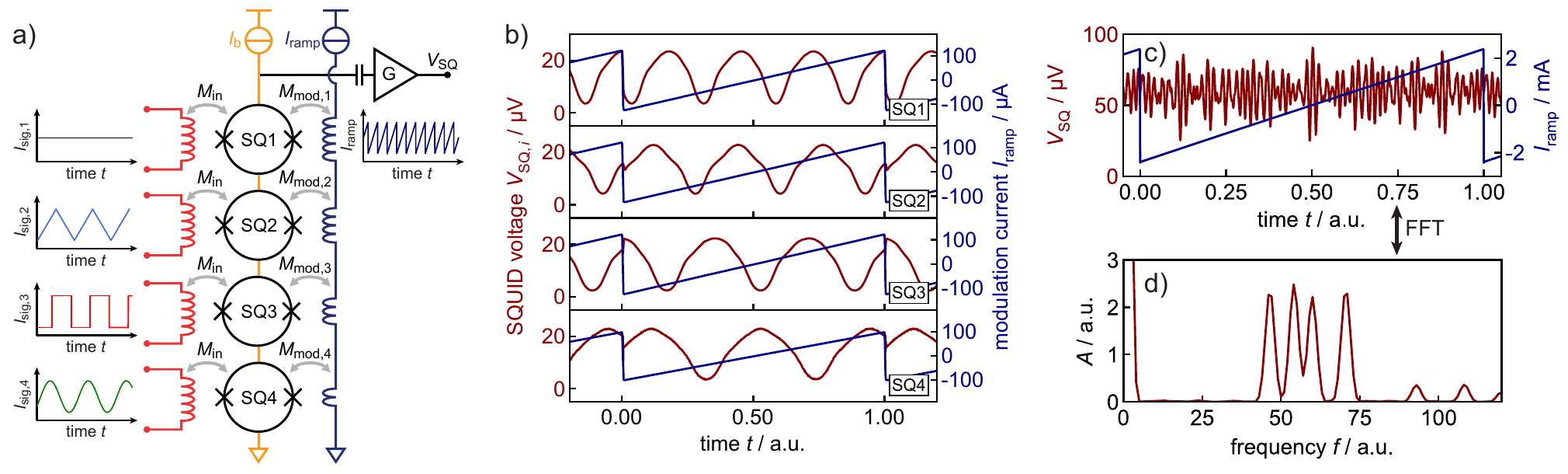}%
    \caption{\label{fig1}(Color online)
    (a) Schematic circuit diagram of a four-channel dc-SQUID multiplexer. It is based on a series array of four current-biased dc-SQUIDs, each being equipped with a tightly coupled input and modulation coil. A periodic, sawtooth-shaped current signal $I_\mathrm{ramp}(t)$ is driven through the common modulation coil and the voltage $V_\mathrm{SQ}$ across the SQUID series array is measured. The currents $I_{\mathrm{sig},i}(t)$ represent exemplary input signals as used for demonstration purposes.
    (b) Modulation current $I_\mathrm{ramp}(t)$ (blue) and resulting output voltage $V_{\mathrm{SQ},i}(t)$ (red) of the individual SQUIDs. For better visibility, we used a rather low amplitude of the flux ramp. In actual experiments, the flux ramp amplitude is much higher and several tenth of periods are induced.
    (c) Output signal $V_\mathrm{SQ}(t)$ of the SQUID array.
    (d) Fast Fourier transform $\mathcal{F}\left[V_\mathrm{SQ}(t)\right]$ of the voltage signal $V_\mathrm{SQ}(t)$ as acquired during one cycle of the flux ramp.}
\end{figure*}

In view of this, we present a MHz frequency-division SQUID multiplexing technique. It is based on flux ramp modulation, a modulation technique that had been originally developed for linearizing the output signal of a microwave SQUID multiplexer \cite{Mat12}. It relies on injecting a periodic, sawtooth-shaped modulation current signal $I_\mathrm{ramp}(t)$ into the modulation coil of a current-sensing SQUID to induce a linearly rising flux ramp with an amplitude of several flux quanta inside the SQUID loop causing the SQUID output voltage to vary according to its flux-to-voltage characteristic. The flux ramp repetition rate sets the effective sampling rate and hence the signal bandwidth. It is chosen such that the input signal appears to be quasi-static within a time frame of the flux ramp. In this configuration, the input signal leads to a constant magnetic flux offset causing a phase-shift of the flux-to-voltage characteristics that depends linearly on the actual amplitude of the input signal. Determining this phase shift, e.g. using Software-defined radio (SDR) based readout electronics, provides an intrinsically linearized measure of the input signal. A phase-shift of $2\pi$ corresponds to a magnetic flux change of one magnetic flux quantum.

Fig.~\ref{fig1}a shows a schematic circuit diagram of a four-channel multiplexer that is based on our multiplexing approach. Four SQUIDs (one SQUID for each readout channel) are connected in series and dc-current biased such that the voltage $V_\mathrm{SQ}$ across the array is the sum of the output voltages $V_{\mathrm{SQ},i}$ of the individual SQUIDs. Each SQUID is equipped with a tightly coupled input coil with mutual inductance $M_{\mathrm{in}}$ that is connected to the actual signal source, e.g. a superconducting pick-up coil or a cryogenic detector. Furthermore, the SQUIDs are equipped with modulation coils, each having a different mutual inductance $M_{\mathrm{mod},i}$ (the choice of values is discussed below), that are serially connected. By injecting a sawtooth-shaped current signal $I_\mathrm{ramp}(t)$ into these coils, a linearly rising flux ramp is induced inside each SQUID loop. The actual amplitude of the flux ramp (magnetic flux signal inside the SQUID loop in units of $\Phi_0$) depends on the mutual inductance $M_{\mathrm{mod},i}$ and the amplitude of the modulation current, the latter being adjusted such that multiple flux quanta are induced inside each SQUID loop. For this reason, the flux ramp causes the output voltage $V_{\mathrm{SQ},i}$ of the $i$-th SQUID to vary according to its flux-to-voltage characteristics, where the number of periods depends on the height of the flux ramp (see Fig.~\ref{fig1}b). The periodic oscillation of the output voltage $V_{\mathrm{SQ},i}(t)$ of the $i$-th SQUID hence acts as a carrier signal which is phase-modulated by the signal source connected to the SQUID input. In case that the mutual inductances $M_{\mathrm{mod},i}$ are properly chosen, the carrier frequencies $f_{\mathrm{c},i} = I_\mathrm{ramp,pp} M_{\mathrm{mod},i} f_\mathrm{ramp} / \Phi_0$ are unique and can be set by the amplitude $I_\mathrm{ramp,pp}$ and repetition rate $f_\mathrm{ramp}$ of the modulation signal. Since each time frame of the flux ramp is used to acquire exactly one sample of the input signal, the flux ramp repetition ramp $f_\mathrm{ramp}$ simultaneously defines the effective sampling rate of the signal and is hence adjusted to the requirements of the specific application. For this reason, the carrier frequencies are in practice mainly set by the amplitude $I_\mathrm{ramp,pp}$ of the modulation signal. The series connection of the individual SQUIDs allows to sum the carriers into the output voltage $V_\mathrm{SQ}$ across the entire array (see Fig.~\ref{fig1}c and d). By channelizing this overall output voltage signal $V_\mathrm{SQ}(t)$ for each cycle of the flux ramp using, for example, digital down converters combined with subsequent low-pass filters, the phase of the individual carriers can be continuously monitored and acquired in real-time. In this sense, $N$ signals can be simultaneously read out using only two bias lines as well as two lines connected to the common modulation coil. However, the number of readout channels that can be simultaneously multiplexed is ultimately limited by wideband SQUID noise which adds on the different carrier signals and whose amplitude increases as $\sqrt{N}$.

\begin{figure*}
    \includegraphics[width=1.0\linewidth]{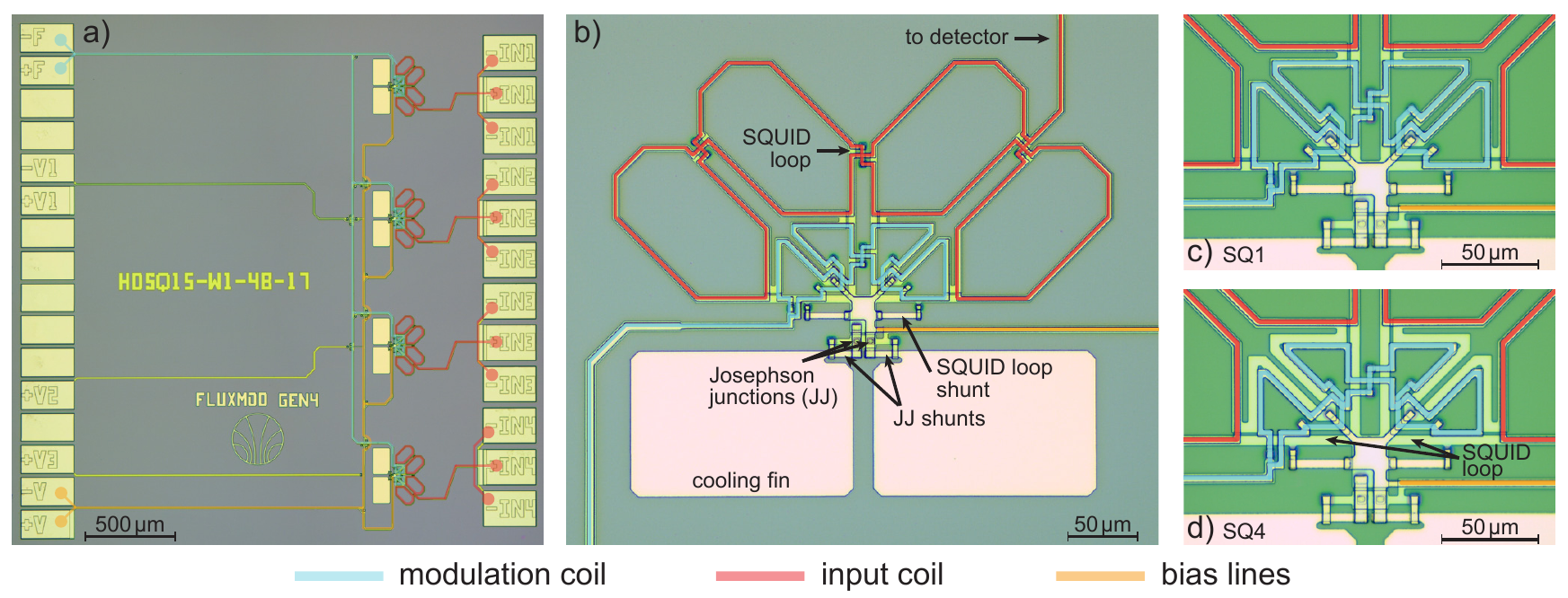}%
    \caption{\label{fig2}(Color online) Optical microscope pictures of the prototype four-channel dc-SQUID multiplexer device presented within this paper. For clarity, the SQUID bias lines, the modulation coil as well as the input coils are colored (compare Fig.~\ref{fig1}) (a) Overview of the entire chip. (b) Magnification of the dc-SQUID ``SQ1''. (c), (d) Magnification of the part of the SQUID loop of the SQUIDs ``SQ1'' and ``SQ4'' to which the modulation coil is inductively coupled.}
\end{figure*}

To demonstrate our multiplexing approach, we designed, fabricated and characterized a four-channel prototype multiplexer. We used our well-established fabrication process for Nb/Al-AlOx/Nb-Josephson tunnel junctions \cite{RN437} as well as a SQUID design that we had previously developed. The SQUID layout is hence not optimized with respect to this multiplexing application. More precisely, the SQUID impedance is not matched to the line impedance, the flux-to-voltage transfer coefficient is not maximized and the on-chip wiring is not optimized for transmitting high-frequency signals. We therefore had to expect a reduced signal to noise ratio. Fig.~\ref{fig2}a shows an optical microscope photograph of one of our fabricated multiplexers. The common modulation coil and the bias lines are colored in blue and orange, whereas the input coils are colored in red. Besides the electrical contact pads that are required for multiplexer operation and that are marked with colored dots, additional pads for initial diagnostics are placed on the left side of the chip. The latter allows, for example, to tap the individual SQUID voltages $V_{\mathrm{SQ},i}$ or to modulate the flux of only a subset of the entire SQUID array. The SQUIDs are parallel gradiometers consisting of four planar, single-turn coils that are connected in parallel (see Fig.~\ref{fig2}b). Each coil is built by two superconducting loops of different size that are connected in series. The bigger loop is used to tightly couple the input coil, while the smaller loop is used to couple the modulation coil. This arrangement allows spatially and thus inductively separating the input and modulation coil to avoid parasitic coupling of the flux ramp into a potential superconducting input circuit as formed, for example, when connecting a superconducting pickup coil to the SQUID input. The Josephson tunnel junctions are located in the lower part of the SQUID and are resistively shunted to ensure a non-hysteretic behavior of the SQUID. The SQUID is equipped with SQUID loop shunts to damp the fundamental SQUID resonance.

For our prototype device, we aimed to equally space the mutual inductances $M_{\mathrm{mod},i}$ in the range $M_{\mathrm{mod,min}}\leq M_{\mathrm{mod},i} < 2M_{\mathrm{mod,min}}$ with $i=1,\ldots,4$ and $M_{\mathrm{mod,min}}$ being the mutual inductance of the weakest coupled SQUID. This prevents higher harmonics of the carrier signals to appear in the target carrier frequency range and ensures that an integer number of flux quanta is induced in each SQUID loop when injecting a proper modulation signal into the common modulation coil. The latter is essential to avoid the occurrence of voltage transients in the flux-to-voltage characteristics that potentially emerge in case that the magnetic flux threading the SQUID loop is different before and after the ramp reset. We set the mutual inductance $M_{\mathrm{mod},i}$ by adjusting the overlap of the modulation coil and the underlying SQUID loop. For the tightest coupled SQUID, i.e. the SQUID with the largest mutual inductance $M_{\mathrm{mod},i}$, the modulation coil runs directly on top of the SQUID loop (see Fig.~\ref{fig2}c) to yield the highest possible magnetic coupling factor. For the other SQUIDs supposed to be weaker coupled, the diameter of the modulation coil is reduced (see, for example, Fig. \ref{fig2}d for the weakest coupled SQUID). We performed numerical inductance calculations of our SQUID design by means of InductEx \cite{RN684} and managed to adjust the overlap to yield almost equally spaced mutual inductance values in the range $29.3\,\mathrm{pH} \leq M_{\mathrm{mod},i} < 58.6\,\mathrm{pH}$. However, the precision of the performed simulations as well as to a minor degree fabrication inaccuracies, e.g. layer thickness inhomogeneities across the wafer as well as alignment errors, led to a slightly non-uniform spacing for the final prototype device. We thus had to deal with voltage transients during the ramp resets deteriorating the phase determination and increasing the white noise level. To avoid this kind of complication for future devices, we plan to use a similar wiring strategy as presently used for code-division multiplexing \cite{RN731}. In particular, we are going to build two-dimensional $M \times N$ multiplexers for which the $N$ modulation coils within a row and the $M$ SQUIDs within a column are serially connected to each other. Assuming that the mutual inductance values within a row are virtually identical, in particular after optimizing the fabrication process, this scheme allows to inject individual flux ramp signals to each column and thus to ensure that each SQUID is modulated with an integer number of flux quanta.

We mounted the fabricated prototype multiplexer on a custom-made sample holder, electrically connected the chip by means of ultrasonic wedge bonds, and immersed the setup enclosed with a soft-magnetic and superconducting shield into a liquid helium transport dewar. We dc-biased the multiplexer using the low-noise bias current source of a commercial high-speed dc-SQUID electronics \footnote{SQUID electronics XXF-1 from Magnicon GmbH, Germany.} and amplified the voltage drop across the array by means of a capacitively coupled, low-cost amplifier cascade \footnote{The amplifier cascade consists of three serially connected amplifiers ZFL-500LN+, ZFL-500+, ZX60-43-S+ for high-frequency signals produced and sold by Mini-Circuits.}. For synthesizing the modulation signal, digitizing the output voltage of the multiplexer as well as real-time channelization and phase determination, we employed a prototype SDR electronics we had previously developed for operating a microwave SQUID multiplexer \cite{RN74, RN355}. The electronics comprises a Xilinx Zynq UltraScale+ FPGA board \footnote{Xilinx Zynq UltraScale+ XCZU9EG-2FFVB1156 MPSoC.} as well as ADC \footnote{ADS54J69 Dual-Channel, 16-Bit, 500-MSPS, Analog-to-Digital Converter. ADS54J69. Texas Instruments.} and DAC \footnote{DAC38J84 Quad-Channel, 16-Bit, 2.5 GSPS, Digital-to-Analog Converter with 12.5 Gbps JESD204B Interface. DAC38J84. Texas Instruments.} evaluation boards. The latter are connected to the FPGA board via FMC connectors and are synchronized via an external $10\,\mathrm{MHz}$ reference. To avoid a degradation of the linear rise of the flux ramp, we replaced the input transformers of the DAC board by inductive transformers with passband of $15\,\mathrm{kHz}$ and $100\,\mathrm{MHz}$. This allowed synthesizing adequate flux ramps signals. However, we later figured out that the effective ac-coupling still distorted the flux ramp linearity manifesting as a change of slope of the ramp during each cycle. The resulting carrier frequency change within each time frame significantly deteriorated the phase determination and ultimately led to an increased white noise floor. For this reason, we plan for dc-coupled DACs to be used in later applications. For channelizing and demodulation of the continuously sampled output voltage signal, we used digital down-conversion with subsequent low-pass filtering for phase estimation.

\begin{figure}[t]
    \includegraphics[width=1.0\linewidth]{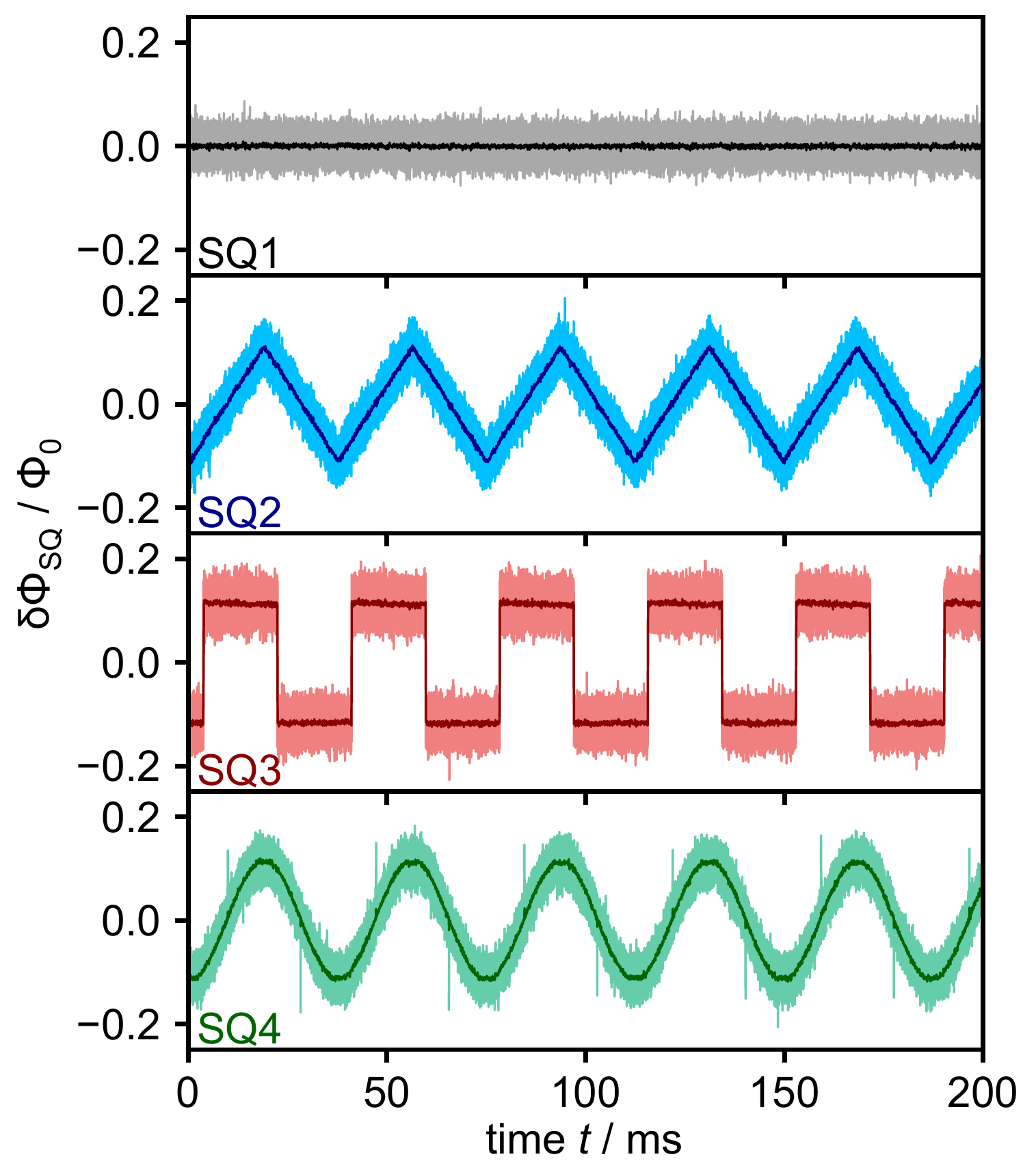}%
    \caption{\label{fig3}(Color online) Magnetic flux excitation $\delta \Phi_{i}$ injected in SQUID ``SQ$i$'' versus time $t$ derived from the demodulation of the output voltage signal of the prototype MHz dc-SQUID multiplexer. The test signals that have been applied to the input coils of the different SQUIDs are illustrated in Fig.~\ref{fig1}(a). The solid lines result from averaging fifty neighboring points to increase visibility.}
\end{figure}

We comprehensively characterized the multiplexer varying, for example, the flux ramp repetition rate $f_\mathrm{ramp}$ or the amplitude $I_\mathrm{ramp,pp}$ of the modulation current $I_\mathrm{ramp}(t)$. The resulting carrier signal frequencies were in the frequency range between $10\,\mathrm{MHz}$ and $50\,\mathrm{MHz}$ and the effective sampling rate was ranging between $200\,\mathrm{kHz}$ and $1.2\,\mathrm{MHz}$. This potentially allows, for example, to read out some tens of high-speed cryogenic microcalorimeters with sub-$\mu$s rise time. Fig.~\ref{fig3} shows as an example the four output signals when connecting test signal generators to the input coils of channels SQ2 to SQ4, running the multiplexer with a flux ramp repetition rate of $f_\mathrm{ramp} = 1\,\mathrm{MHz}$ and adjusting the amplitude to yield carrier frequencies of about $26.6\,\mathrm{MHz}$, $32.5\,\mathrm{MHz}$, $38.0\,\mathrm{MHz}$, and $45.5\,\mathrm{MHz}$. The input signals are clearly resolved. This proves that our multiplexing technique is performing as intended. The cross-talk between different channels was measured by applying a sinusoidal test signal to one channel and measuring the spectral intensity of the associated frequency bin in the other channels. For all channels, the crosstalk is below $-40\,\mathrm{dB}$. The white noise level of all four channels is about $25\,\mu\Phi_0/\sqrt{\mathrm{Hz}}$ and therefore seems to be rather high. However, several sources contribute to this elevated noise level whose influence be minimized for future devices by optimizing the overall setup. This includes the small SQUID flux-to-voltage transfer coefficient, the voltage noise of the used amplifier cascade, the noise penalty due to flux ramp modulation \cite{Mat12}, the remaining non-linearity of the flux ramp, the impedance mismatch as well as the non-optimized device design.

In conclusion, we demonstrated a MHz frequency-division dc-SQUID multiplexing technique. We showed that our approach avoids some drawbacks of other existing multiplexing techniques and allows for reading out signal sources with MHz signal bandwidth at cryogenic temperatures. In combination with the large dynamic range\footnote{In theory the dynamic range is infinitely high. In practice, the dynamic range might be limited by the white noise level, the mutual inductance between the input coil and the SQUID loop and the ampacity of the input coil.} that is comparable or even higher than the dynamic range of standard setups using flux-locked loop and that results from converting the input signal into a phase shift, our multiplexing approach paves the way for realizing various applications requiring a simple setup, large signal bandwidth and dynamic range as well as low noise.

\begin{section}{Data availability}
The data that support the findings of this study are available from the corresponding author upon reasonable request.
\end{section}

\begin{acknowledgments}
We would like to thank T. Wolf as well as cleanroom team of the Kirchhoff-Institute for Physics for their great technical support during device fabrication. The work was performed in the framework of the DFG research unit FOR 2202 (funding under grant EN299/7-2). The research leading to these results has also received funding from the European Union’s Horizon 2020 Research and Innovation Programme, under Grant Agreement No 824109. Furthermore, we greatly acknowledge funding by the German Federal Ministry of Education and Research (funding under grant BMBF FKZ 05P19VHFA1).
\end{acknowledgments}

\bibliography{literature}

\end{document}